\documentclass[
twocolumn,
pre,
aps,
superscriptaddress,
showpacs,
preprintnumbers,
amsmath,
amssymb]{revtex4-1}
\usepackage{bm}

\usepackage[dvipdfmx]{graphicx}
\usepackage[dvipdfmx,usenames]{color}

\begin{document}

\title{Mechanical classification of voice quality}

\author{Akito Yoshida}
\email{a-yoshida@isi.imi.i.u-tokyo.ac.jp}
\affiliation{Department of Mechano-Informatics, University of Tokyo, Tokyo 113-8656, Japan}
\author{Shigeru Shinomoto}
\email{shinomoto.shigeru.6e@kyoto-u.ac.jp}
\affiliation{Department of Physics, Kyoto University, Kyoto 606-8502, Japan}
\affiliation{Brain Information Communication Research Laboratory Group, ATR Institute International, Kyoto 619-0288, Japan}

\begin{abstract}
While there is no \textit{a priori} definition of good singing voices, we tend to make consistent evaluations of the quality of singing almost instantaneously. Such an instantaneous evaluation might be based on the sound spectrum that can be perceived in a short time. Here we devise a Bayesian algorithm that learns to evaluate the choral proficiency, musical scale, and gender of individual singers using the sound spectra of singing voices. In particular, the classification is performed on a set of sound spectral intensities, whose frequencies are selected by minimizing the Bayes risk. This optimization allows the algorithm to capture sound frequencies that are essential for each discrimination task, resulting in a good assessment performance. Experimental results revealed that a sound duration of about 0.1 sec is sufficient for determining the choral proficiency and gender of a singer. With a program constructed on this algorithm, everyone can evaluate choral voices of others and perform private vocal exercises.

\end{abstract}

\maketitle

\section{Introduction}

A professional assessment of singing performances is based on many criteria such as musical pitch, rhythm, breathing, and vibrato. However, a lay audience often makes instantaneous and consistent judgment of the overall quality of singing without evaluating these criteria. For instance, an audience of a musical contest can react to every new singing in a few seconds. Such an instantaneous evaluation might be based on the sound spectrum that can be perceived in a short time \cite{Bartholomew1934}. In particular, it has been reported that the sound spectra of well-trained male opera singers exhibit a significant peak at about 3 kHz, which is absent in untrained singers \cite{Sundberg1990}. This spectral peak is called the singer's formant; it has been analyzed with spectral modeling synthesis and has been issued as the singing power ratio \cite{Omori1996,Watts2006,Aithal2011,Usha2017} or quality ratio \cite{Manfredi2015}. 

Here we considered the sound spectral information and developed a Bayesian algorithm that learns choral proficiency from sample singing voices. We also let the algorithm identify the musical scale and whether the owner of the voice is male or female. The factors useful for each categorization were found by optimizing the model performance. We also estimated the minimum duration of a sound required for determining the considered characteristics of individual singers. By embedding the algorithm into an application program, everyone may evaluate the choral voice and perform private vocal exercises by oneself, similar to other applications that evaluate karaoke singing by taking into account the vocal pitch accuracy \cite{Welch1989,Hoppe2006,Tsai2012,Goto2014} and vibrato \cite{Nakano2006}.

For this purpose, we collected singing voices of subjects as training examples and categorized them according to whether the subjects had participated in choral groups (referred to as singers) or not (referred to as non-singers), and let the algorithm learn this category. Since the participation in choral groups does not necessarily guarantee good voices, we asked a professional voice trainer to evaluate the recorded voices and confirmed that our classification was strongly correlated with the choral voice quality. The proposed algorithm was trained and evaluated using cross-validation.

\section{Methods} 

\subsection{Subjects}

We recruited 50 subjects from undergraduate and graduate courses at Kyoto University (ages ranged from 19 to 27). We classified them into singers and non-singers according to whether or not they had participated in a mixed chorus club more than one year. As a result, we acquired singing voices from 23 singers (11 males and 12 females) and 27 non-singers (14 males and 13 females). We asked a professional voice trainer to evaluate their recorded voices. 

\subsection{Recording}

\begin{figure}[htbp]
\includegraphics[width=8.6cm]{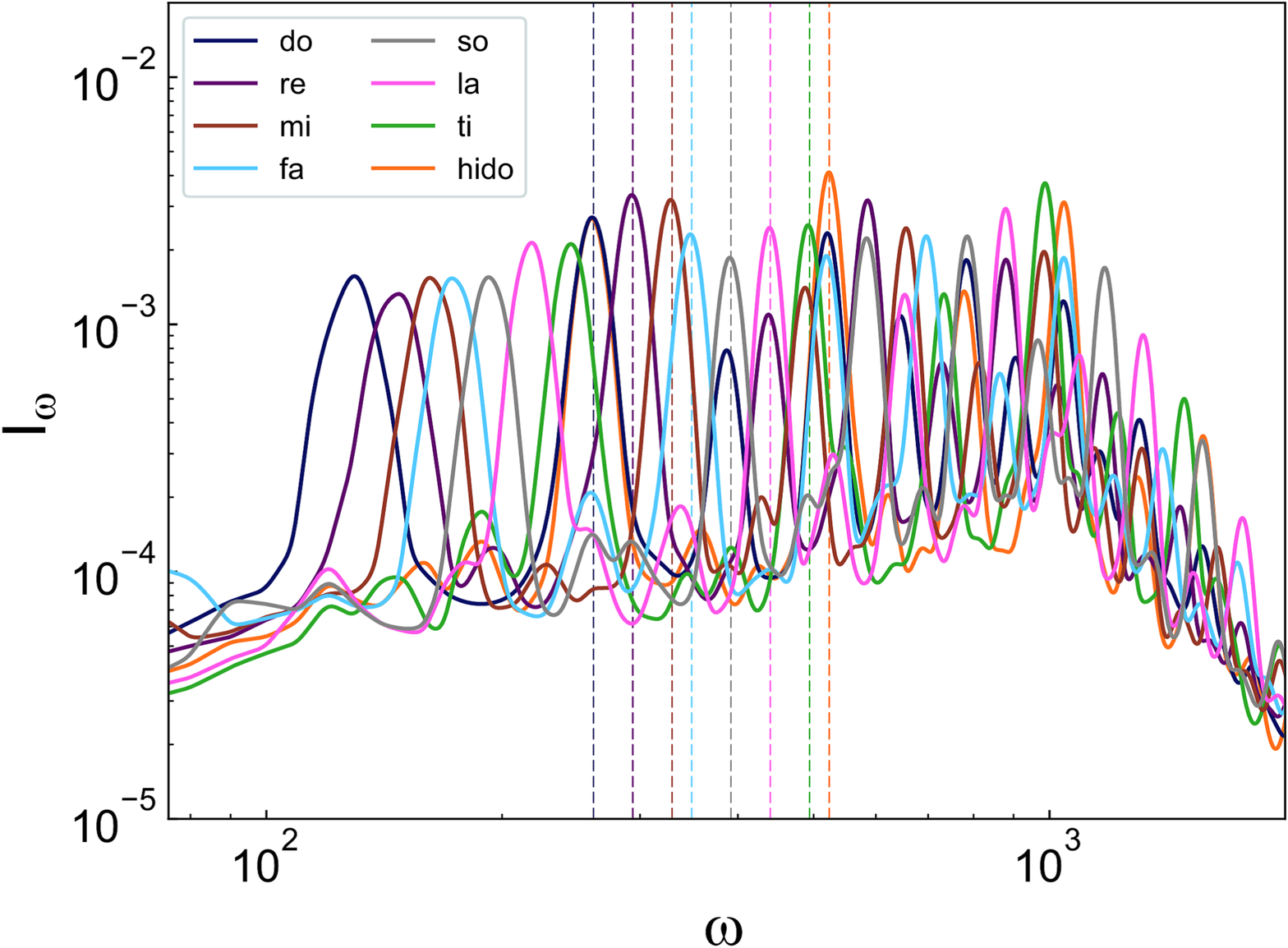}
\caption{Average power spectra of singers' voices at every musical scale of ``do re mi fa so la ti do'' in C major. The vertical axis represents the normalized power spectra $I_{\omega}$, while the horizontal axis represents the frequency $\omega$. The vertical dashed lines represent the basic frequencies of the eight musical scales.}
\label{musicalscales}
\end{figure}
\begin{figure}[htbp]
\includegraphics[width=8.6cm]{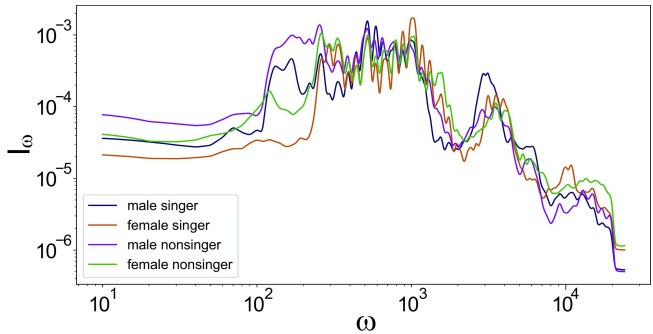}
\caption{Average power spectra of the four categories defined as combinations of singer or non-singer and male or female.}
\label{fourcategories}
\end{figure}

Each subject was asked to sing ``a'' sound for 1.5 sec by following a piano at every musical scale of ``do re mi fa so la ti do'' in C major. The voices were recorded using a MacBook Air at a sampling rate of 48,000 Hz. We divided the sound data into intervals of $\Delta=0.1$ sec each and selected 10 pieces with high intensity. For each piece of sound data, the power spectrum or the amplitude spectrum is obtained with a fast Fourier transform (FFT). To exclude the effect of the difference in the sound intensity, we normalized the original power spectrum $I^{0}_{\omega}$ as
\begin{equation}
I_{\omega} =\frac{I^{0}_{\omega}}{\int d \omega I^{0}_{\omega}}.
\label{normalizedpower}
\end{equation}
We use the normalized power spectrum $I_{\omega}$ throughout the analysis. 

Figure~\ref{musicalscales} depicts the normalized power spectra of eight musical scales, each averaged over the singers' voices. Figure~\ref{fourcategories} shows four power spectra classified according to the combinations of singer or non-singer and male or female.

\subsection{Inferring voice categories from sound spectrum}

We developed a Bayesian method for assessing voice categories (singer or non-singer and male or female) from a sound spectrum. This can be achieved by obtaining the conditional distribution $P(c|\{I_{\omega}\})$ of a category or class $c$ given a sound spectrum $\{I_{\omega}\}$, which is a function of the frequency $\{\omega\}$. In a practical setting, we divide the frequency range from 0 to 20 kHz into intervals of 10 Hz. The sound spectrum comprised 2,000 elements. To make the problem tractable, we selected a small number of power spectral values $\mathbf {x} \equiv$ $\{ I_{\omega_1}$, $I_{\omega_2}$, $\cdots$, $I_{\omega_D} \}$ taken at frequencies of $\{ \omega_1$, $\omega_2$, $\cdots$, $\omega_D \}$ and represented the conditional distribution $P(c|\{I_{\omega}\})$ as $P(c|\mathbf {x})$. The frequencies were selected using a modified version of the method for detecting coins from the colliding sound introduced by one of the authors~\cite{Furukawa2018}. The difference from the original method are summarized below.

The conditional distribution $P(c|\mathbf {x})$ was obtained with Bayes' theorem:
\begin{eqnarray}
P(c|\mathbf {x})= \frac{P(\mathbf {x}|c) P(c)}{\sum_{c'} P(\mathbf {x}|c') P(c')},
\label{bayes}
\end{eqnarray}
where $P(c)$ denotes the prior distribution of the class $c$. Here we took the equal prior, $P(c)=1/C$, where $C$ denotes the number of categories (for instance, $C=2$ for singer vs non-singer). We assumed the class-conditional density $P(\mathbf {x}|c)$ used in the framework of Gaussian discriminant analysis~\cite{Bishop2006,Murphy2012}:
\begin{eqnarray}
P(\mathbf {x}|c)= \frac{\exp \left\{ - \frac{1}{2}(\mathbf {x}-\boldsymbol {\mu}_c)^{\mathrm{T}} \bm {\Sigma}_c ^{-1}(\mathbf {x}-\boldsymbol {\mu}_c) \right\}}{(2 \pi)^{D/2} |\bm {\Sigma}_c |^{1/2}},
\label{classconditional}
\end{eqnarray}
where $\boldsymbol{\mu_c}$ and $\bm{\Sigma_c}$ denote the mean and covariance estimated from the training data, respectively. 

For each sound spectrum obtained from the sound data in an interval $\Delta =0.1$ sec, we estimated the conditional distribution $P(c|\mathbf {x})$ using Eqs. (\ref{bayes}) and (\ref{classconditional}). While each time bin of $\Delta = 0.1$ sec gives a fair inference, integrating the knowledge over 10 intervals may strengthen the inference. Hence, we used the arithmetic mean of the posterior distributions $P(c|\mathbf {x})$ computed for 10 intervals. Using the average conditional distribution, we obtained the maximum {\it a posteriori} (MAP) inference as
\begin{eqnarray} 
c_{\mathbf {x}} = \arg \max_{c} P(c|\mathbf {x}).
\end{eqnarray}

Given the number of power spectra or dimension $D$, we selected frequencies $\{ \omega_1$, $\omega_2$, $\cdots$, $\omega_D \}$ to make the inference as efficient as possible. This can be achieved by choosing the frequencies that let the class-conditional densities $\{ P(\mathbf {x}|c)\}_c$ separate maximally, so that the membership of each sound is optimally distinguished. As a guiding principle, we suggest minimizing the Bayes risk $R$ or maximizing a theoretical performance $1-R$ as follows.
\begin{eqnarray}
1-R=\int d\mathbf {x} \, P(c_{\mathbf {x}}|\mathbf {x})P(\mathbf {x})=\int d\mathbf {x} \, P(\mathbf {x}|c_{\mathbf {x}})P(c_{\mathbf {x}}).
\label{bayesrisk}
\end{eqnarray}

We explored a set of frequencies $\{ \omega_1$, $\omega_2$, $\cdots$, $\omega_D \}$ by readjusting every single frequency $\omega_i$ to minimize the Bayes risk $R$ sequentially from $i=1$ to $D$, similar to Algorithm 1 in \cite{Furukawa2018}. In the present study, we searched the space of $\log \omega$ instead of real $\omega$ because the important frequencies are distributed in the range lower than that of inferring coins. To perform the search in the logarithmic scale, we re-sampled the power spectrum in the $\log \omega$ axis by interpolating $\log I_{\omega}$.

\section{Results} 

\subsection{Inferring musical scales}

\begin{figure}[htbp]
\includegraphics[width=8.6cm]{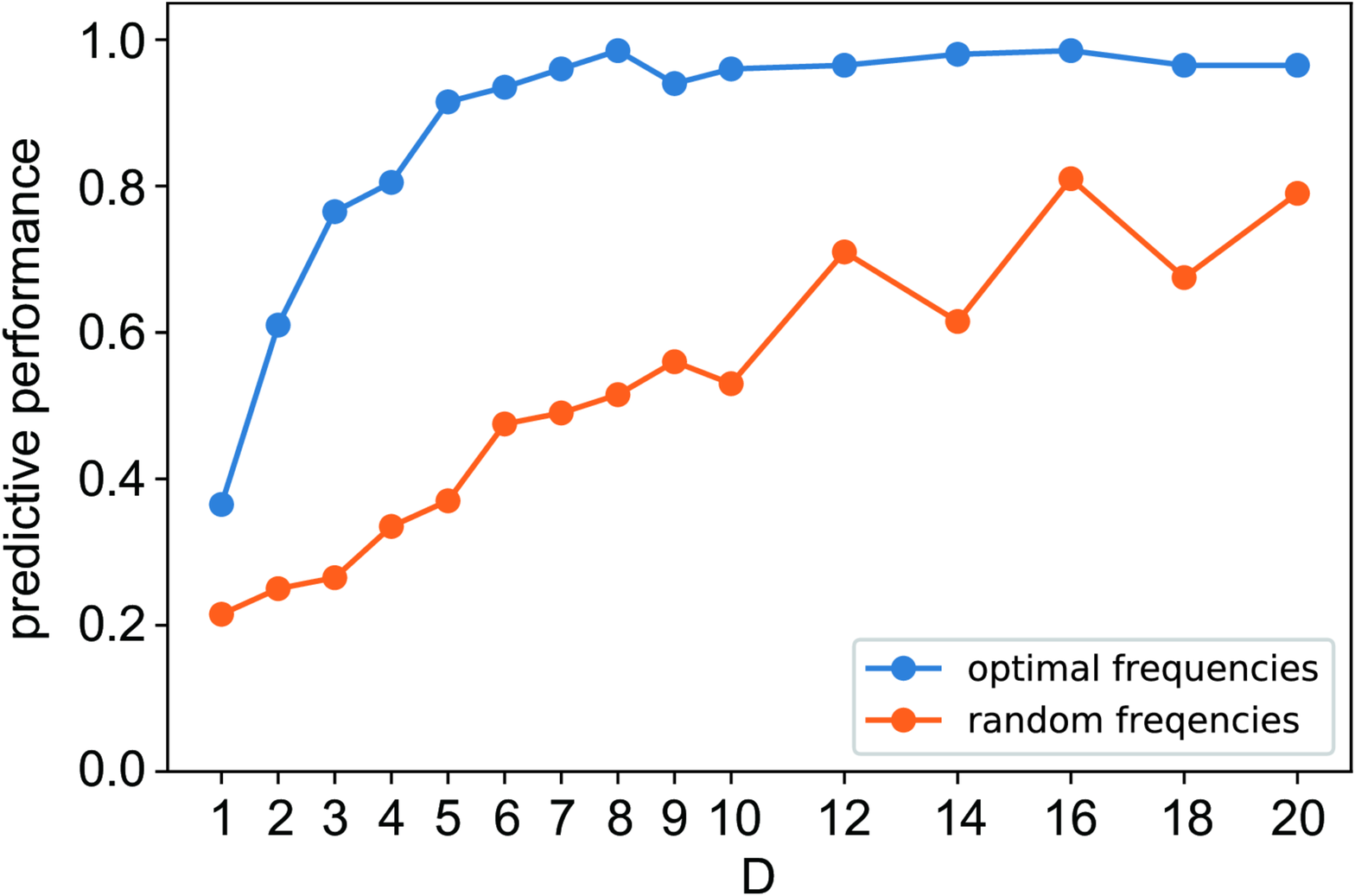}
\caption{Cross-validated performance of the proposed algorithm in identifying musical scales. The performance achieved by selecting frequencies using Bayes risk minimization is plotted against the performance obtained using randomly chosen frequencies.}
\label{scaleperformance}
\end{figure}
\begin{figure}[htbp]
\includegraphics[width=8.6cm]{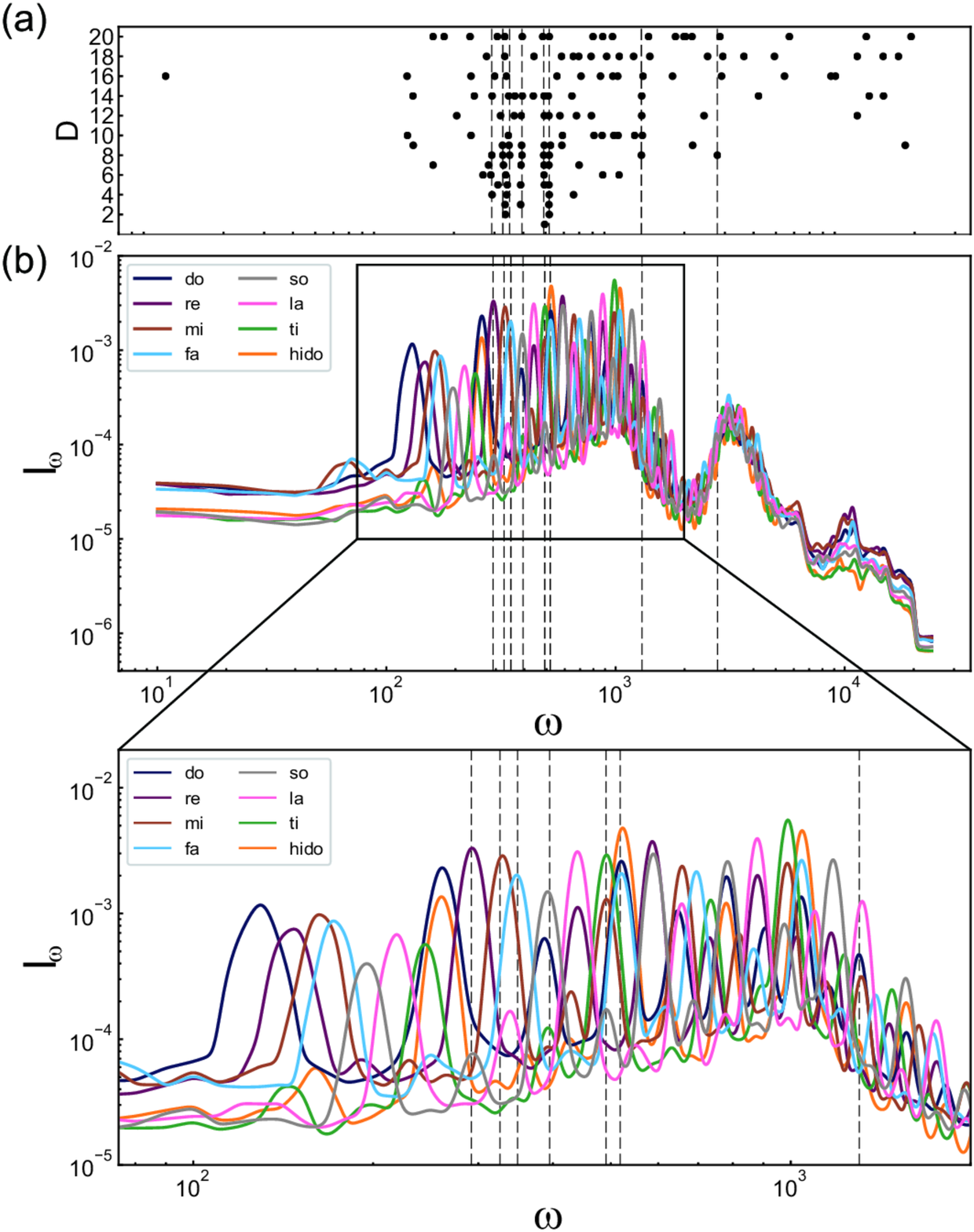}
\caption{Selected frequencies and power spectra. (a) Frequencies selected by minimizing the Bayes risk for each dimensionality $D$. (b) Power spectra of eight musical scale voices averaged over all singers. Dashed lines indicate the selected frequencies for $D=8$.}
\label{scalefrequencies}
\end{figure}

First, we used the proposed algorithm to infer the musical scale of a voice, assigning it to match one of the ``do re mi fa so la ti do'' in C major. In this experiment, we used voices of 23 singers, who are expected to sing on the key. The predictive performance was estimated using cross-validation; we repeat the testing over the voices of randomly chosen 5 subjects using the algorithm trained on the remaining 18 subjects.

We found that the performance of identifying musical scales increases on average with the number of parameters or dimension $D$. It can be noticed from Fig.~\ref{scaleperformance} that the performance is much higher for the case of selecting frequencies $\{ \omega_1$, $\omega_2$, $\cdots$, $\omega_D \}$ by minimizing the Bayes risk compared to the case of randomly selected frequencies from the uniform distribution in $\log \omega$. 

Figure \ref{scalefrequencies}(a) shows the frequencies $\{ \hat\omega_1$, $\hat\omega_2$, $\cdots$, $\hat\omega_D \}$ selected for the entire data set by minimizing the Bayes risk $R$ for various dimensions $D$. The frequencies selected for different dimensions $D$ tend to be similar to each other. By plotting the power spectra of the eight musical scales, we can see that the selected frequencies mostly coincide with the basic frequencies of the musical scales (Fig.~\ref{scalefrequencies}(b)). This indicates that frequencies can be selected reasonably by simply following the principle of the Bayes risk minimization.

\subsection{Inferring male or female}

\begin{figure}[htbp]
\centering
\includegraphics[width=8.6cm]{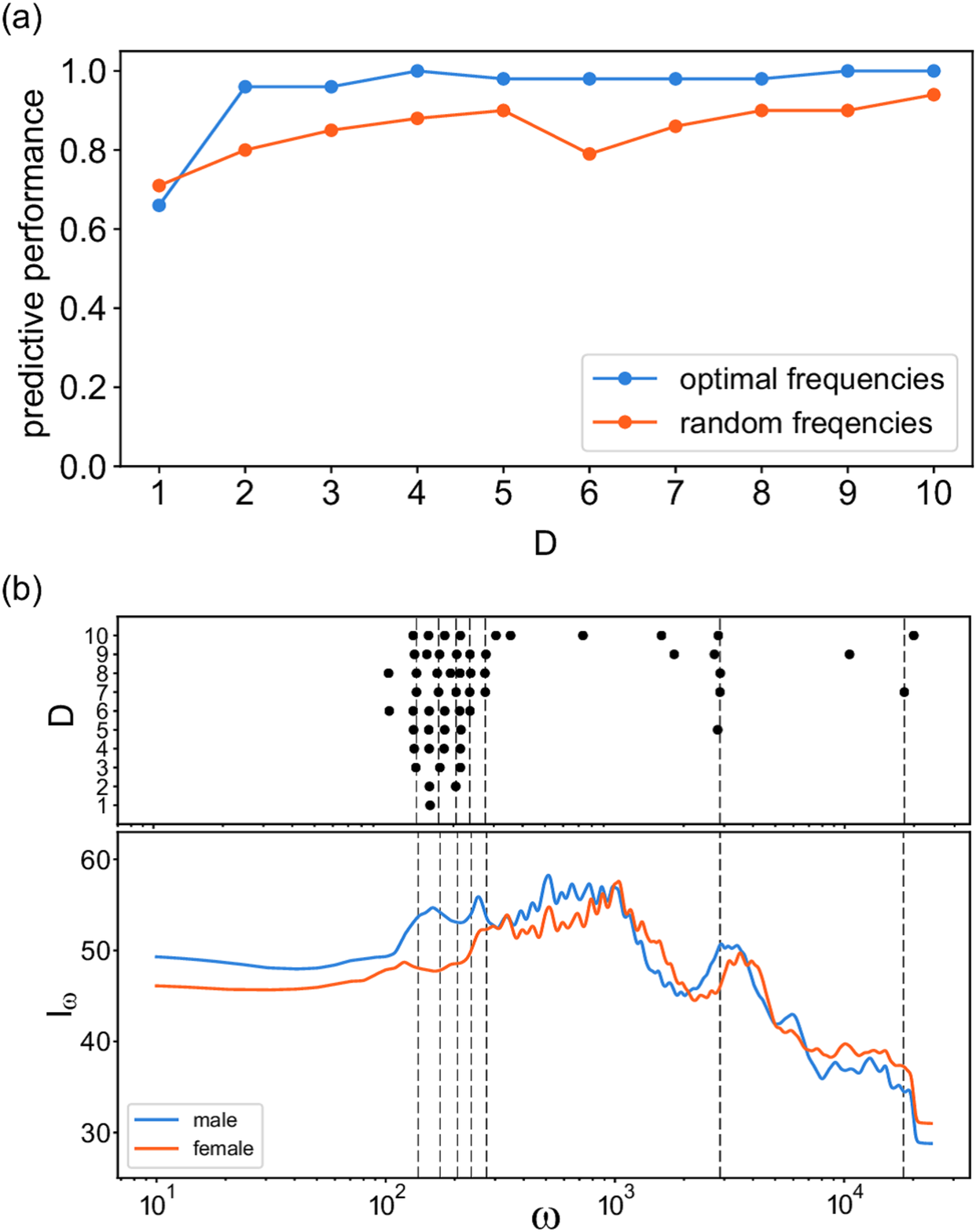}
\caption{
(a) Performances of the algorithm in distinguishing between males and females based on their voices.
(b) Selected frequencies and power spectra. Frequencies selected by minimizing the Bayes risk, and the averaged power spectra for males and females.
}
\label{MF}
\end{figure}

Next, we used the algorithm to identify the gender of singers from their voices. The predictive performance was estimated using cross-validation; we repeat the testing of 5 male and 5 female voices randomly chosen from 25 males and 25 females, using the algorithm trained with the remaining 40 subjects.

The performance of identifying the gender of singers was almost perfect for $D \ge 2$, when using the optimized frequencies (Fig.~\ref{MF}(a)). The frequencies selected by minimizing the Bayes risk mostly ranged from 100 to 250 Hz (Fig.~\ref{MF}(b)), implying that the algorithm made use of the differences in the power in the lower tones; males were producing lower harmonics for each given key compared to females.

\subsection{Inferring singer or non-singer}

\begin{figure}[htbp]
\centering
\includegraphics[width=8.6cm]{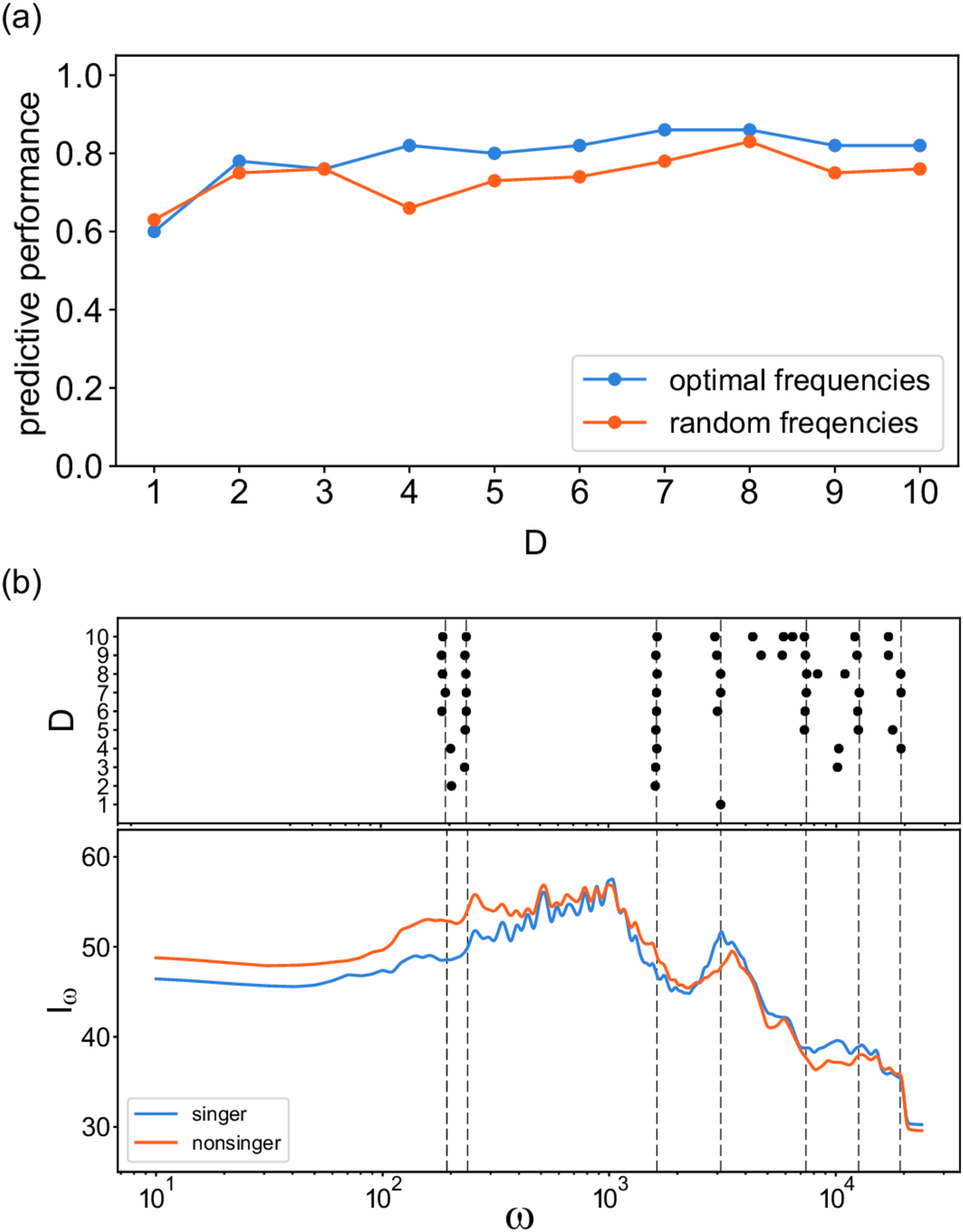}
\caption{
(a) Performances of the algorithm in discriminating singers and non-singers.
(b) Selected frequencies and power spectra. Frequencies selected by minimizing the Bayes risk and the averaged power spectra for males and females. 
}
\label{SN}
\end{figure}
\begin{figure}[htbp]
\centering
\includegraphics[width=8.6cm]{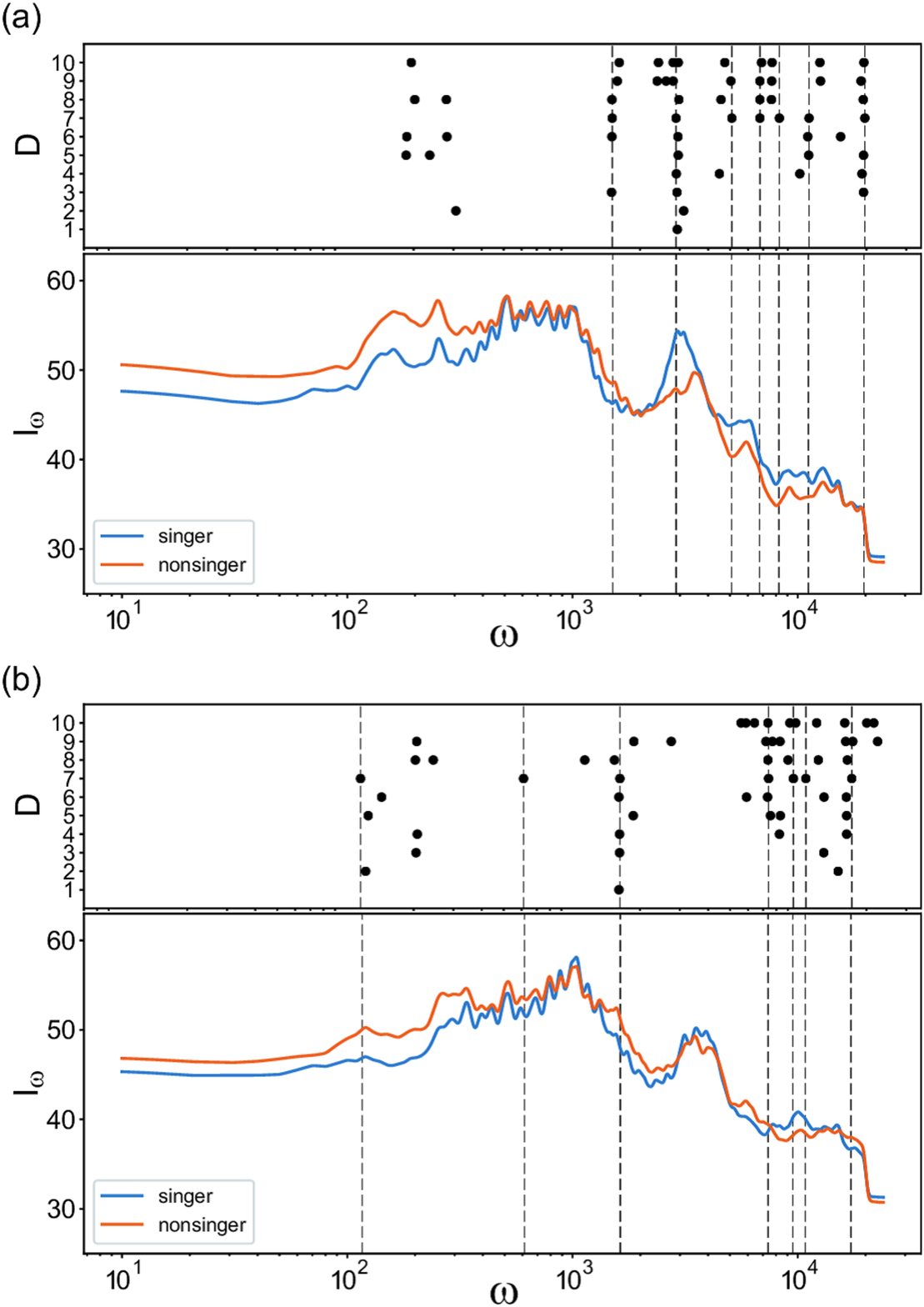}
\caption{
(a) Power spectra of singers and non-singers of males and selected frequencies.
(b) Power spectra of singers and non-singers of females and selected frequencies.
}
\label{MFsingerfrequencies}
\end{figure}

The predictive performance of the algorithm for distinguishing between singers and non-singers was estimated using cross-validation. In particular, we repeated the testing of 5 singer and 5 non-singer voices randomly chosen from 23 singers and 27 non-singers, respectively, using the algorithm trained with the remaining 40 subjects.

The predictive performance was improved by selecting frequencies and became more than 80\% correct (Fig.~\ref{SN}(a)). Several specific frequencies were selected by minimizing the Bayes risk (Fig.~\ref{SN}(b)), including the singer's formant at about 3 kHz.

The predictive performance of distinguishing between singers and non-singers became higher (about 90 \%) when we limited the population to only males. Namely, we estimated the model performance by testing voices of 3 singers and 3 non-singers randomly chosen from 11 male singers and 14 male non-singers, using the algorithm trained with the remaining 19 subjects. In this case, the singer's formant at about 3 kHz was more prominent (Fig.~\ref{MFsingerfrequencies}(a)). In contrast, the predictive performance of distinguishing between singers and non-singers was lower (about 70 \%) when we limited the population to only females. In this case, there was no prominent singer's formant at 3kHz; interestingly however, there was a small peak at about 10 kHz (Fig.~\ref{MFsingerfrequencies}(b)).

\subsection{Inferring male or female and singer or non-singer}

\begin{figure}[htbp]
\centering
\includegraphics[width=8.6cm]{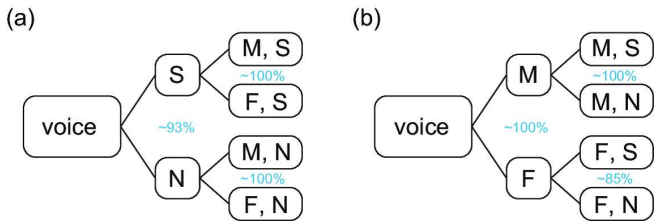}
\caption{
Inferring singer or non-singer (S/N) and male or female (M/F) in different order.
(a) Infer S/N and then infer M/F, $P(\textrm{M/F} | \textrm{S/N}, \mathbf {x})P(\textrm{S/N} | \mathbf {x})$.
(b) Infer M/F and then infer S/N, $P(\textrm{S/N} | \textrm{M/F}, \mathbf {x})P(\textrm{M/F} | \mathbf {x})$.
}
\label{order}
\end{figure}
\begin{figure}[htbp]
\includegraphics[width=8.6cm]{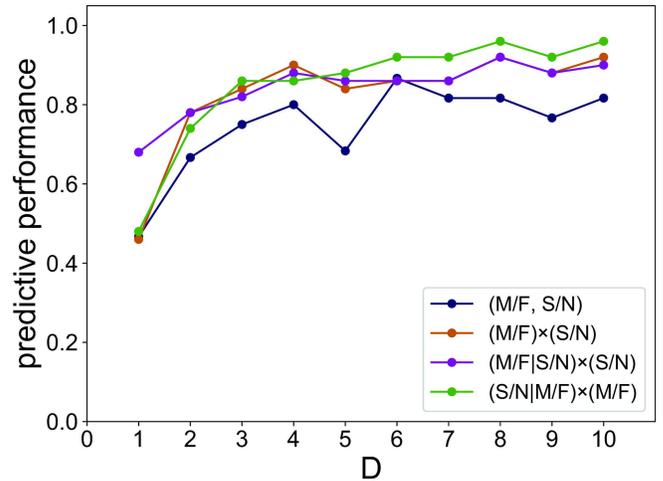}
\caption{Performance of the four methods for capturing M/F and S/N categories.}
\label{orderperformance}
\end{figure}

There are four ways of capturing two types of information (male or female (M/F) and singer or non-singer (S/N)) from the sound spectrum or $\mathbf {x} \equiv$ $\{ I_{\omega_1}$, $I_{\omega_2}$, $\cdots$, $I_{\omega_D} \}$ as described below.

\begin{itemize}
\item Infer M/F and S/N independently: \[P(\textrm{M/F} | \mathbf {x}) P(\textrm{S/N} | \mathbf {x})\]
\item Infer S/N and then infer M/F (Fig.~\ref{order}(a)): \[P(\textrm{M/F} | \textrm{S/N}, \mathbf {x})P(\textrm{S/N} | \mathbf {x})\]
\item Infer M/F and then infer S/N (Fig.~\ref{order}(b)): \[P(\textrm{S/N} | \textrm{M/F}, \mathbf {x})P(\textrm{M/F} | \mathbf {x})\]
\item Infer M/F and S/N simultaneously: \[P(\textrm{M/F}, \textrm{S/N} | \mathbf {x})\]
\end{itemize}

We compared the performance of these four types of inference in Fig.~\ref{orderperformance}. Here, a set of frequencies is optimized in each dimensionality $D$. The performance for the case of inferring M/F and S/N simultaneously is slightly lower compared to that of the other three methods.

\subsection{Comparison between the evaluations made by the algorithm and a professional voice trainer}

\begin{figure*}[htbp]
\centering
\includegraphics[width=14cm]{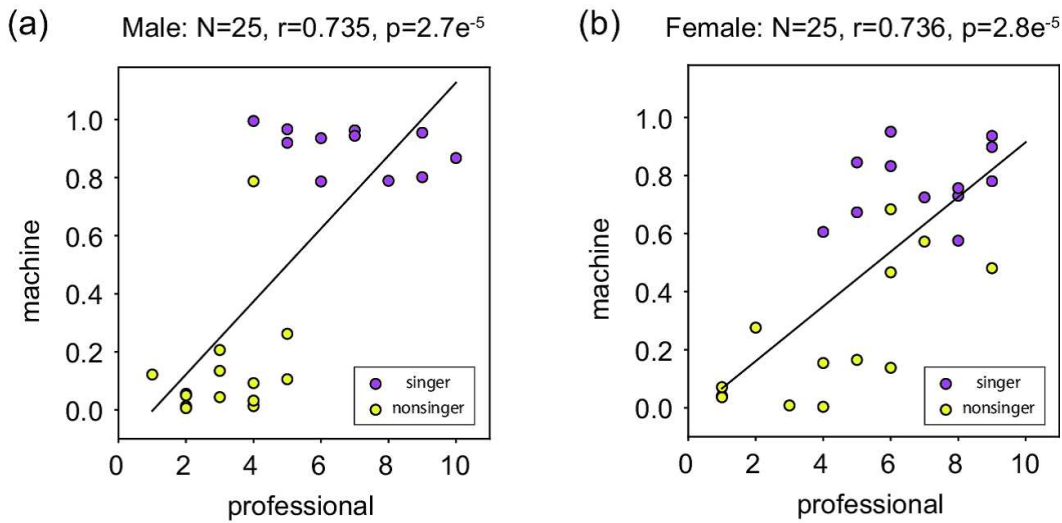}
\caption{
Comparison between the evaluations made by the trained algorithm and a professional voice trainer. The vertical axis represents the posterior distribution of the Bayesian estimator for the singer or non-singer categories $P(\textrm{S/N} | \mathbf {x})$. 
}
\label{professional}
\end{figure*}

Since the participation in a choral group does not necessarily guarantee good voices, we asked a professional voice trainer to evaluate the recorded voices. By comparing the posterior distribution of the Bayesian estimator $P(\textrm{S/N} | \mathbf {x})$ with the evaluation score provided by the professional voice trainer, we confirmed that they are strongly correlated (Fig.~\ref{professional}).

\subsection{Duration of a sound required for the estimation}

\begin{figure*}[htbp]
\centering
\includegraphics[width=14cm]{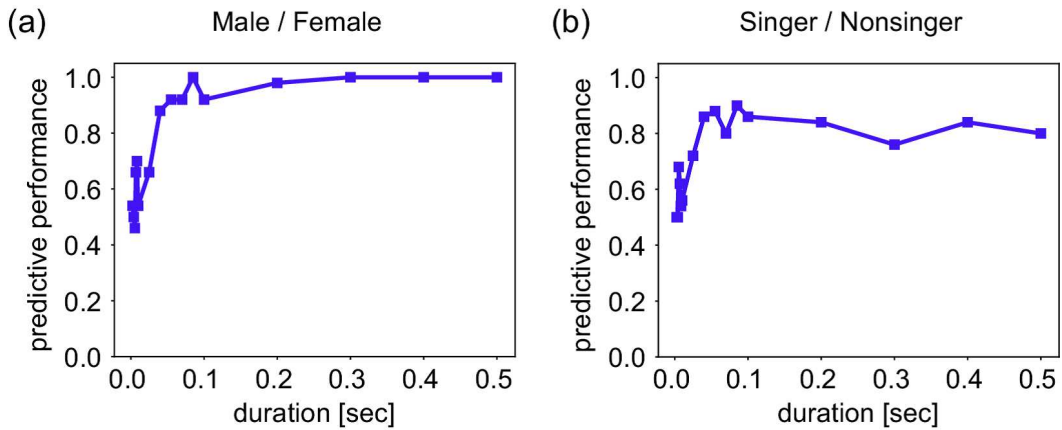}
\caption{
Performance of inference vs sound length. 
(a) Inference of the male and female (M/F) categories. 
(b) Inference of the singer and non-singer (S/N) categories.
}
\label{duration}
\end{figure*}

To estimate the minimum duration of a sound required for determining characteristics such as male or female and singer or non-singer, we estimated the predictive performance of the algorithm by shortening the duration of analysis $10 \times \Delta$ from 1 sec. It can be observed from Fig.~\ref{duration} that the performance is mostly stable at $10 \times \Delta = 0.1$ sec, implying that the inference of male or female and evaluation of singing voices can be performed almost instantaneously. 

\section{Discussion}

The professional assessment of singing is a complicated task accounting for various factors such as musical pitch, rhythm, breathing, and vibrato. There have been attempts to achieve the automatic evaluation of singing by considering these factors. For instance, Nakano et al. presented an automated algorithm for evaluating the singing skill by detecting the pitch interval accuracy and vibrato~\cite{Nakano2006}. There are studies utilizing neural network learning that aim to achieve high performance in the assessment~\cite{Bohm2017,Vidwans2017,Pati2018,Wu2018,Zhang2019}.

In the present study, we have developed an algorithm for assessing music performances based solely on the sound spectrum of singing voices. The algorithm can evaluate the choral proficiency of a singer fairly well without accounting for the time variability of songs. In particular, our algorithm was able to make a reasonable assessment from a short passage of a singing voice of about 0.1 sec, with the information of rhythm, breathing, and vibrato being unavailable. This is similar to audiences making instantaneous voice evaluations when hearing new singing performances in musical contests.

For each task of the categorization regarding the choral proficiency, musical scale, and gender of a singer, our learning algorithm selects frequencies of the sound spectrum by minimizing the Bayes risk. This optimization allows the algorithm to capture sound frequencies that are essential from each discrimination task, resulting in a good assessment performance. In particular, frequencies close to the basic frequencies or their harmonics were selected to separate musical scales; low frequencies of 100-300 Hz, at which the male voices have stronger power, were selected for separating males and females; and frequencies of about 3 kHz and 10 kHz were selected for males and females, respectively, when separating singers and non-singers. The good performance achieved in distinguishing between male singers and non-singers may be largely due to the detection of singer's formant at 3 kHz. The good performance in categorizing female singers and non-singers may be largely due to the peak at 10 kHz. It should be noted that the categorization of the female singing proficiency was not as good as that achieved for males. Regarding the duration of hearing a sound, it was established that 0.1 sec is enough to distinguish between males and females or singers and non-singers.

When analyzing the singing proficiency, we found the singer's formant or prominent peak of the power spectrum to be at 3 kHz for male singers and 10 kHz for female singers. We were interested to find out whether these peaks are the cause of good singing voices or epiphenomena of good voices. To investigate this, we generated colored noises with frequencies distributed at about 3kHz or 10 kHz, added them into the singing voices, and asked the professional voice trainer to evaluate the sounds. As a result, we did not observe significant changes in the evaluation of good voices. Nevertheless, the current experimental setup was not perfect for answering this question and we plan to investigate it in more detail in the future. 

In the present study, we focused on proficiency in classical choral singing. Since the method of Bayesian learning presented in this study is a general framework applicable to any sounds, it would be interesting to apply the method to other styles of singing, such as pop, soul, country, folk, and metal, and see the difference in the important frequencies for evaluating the proficiency of singing in these styles. 

\section*{ACKNOWLEDGMENTS}
We thank Kazuki Nakamura and Sou Shinomoto for collecting voices, and Kinya Nakanishi for evaluating singing voices. S.S. is supported by JST CREST Grant Number JPMJCR1304, and the New Energy and Industrial Technology Development Organization (NEDO).


\begin{thebibliography}{20}%
\makeatletter
\providecommand \@ifxundefined [1]{%
 \@ifx{#1\undefined}
}%
\providecommand \@ifnum [1]{%
 \ifnum #1\expandafter \@firstoftwo
 \else \expandafter \@secondoftwo
 \fi
}%
\providecommand \@ifx [1]{%
 \ifx #1\expandafter \@firstoftwo
 \else \expandafter \@secondoftwo
 \fi
}%
\providecommand \natexlab [1]{#1}%
\providecommand \enquote  [1]{``#1''}%
\providecommand \bibnamefont  [1]{#1}%
\providecommand \bibfnamefont [1]{#1}%
\providecommand \citenamefont [1]{#1}%
\providecommand \href@noop [0]{\@secondoftwo}%
\providecommand \href [0]{\begingroup \@sanitize@url \@href}%
\providecommand \@href[1]{\@@startlink{#1}\@@href}%
\providecommand \@@href[1]{\endgroup#1\@@endlink}%
\providecommand \@sanitize@url [0]{\catcode `\\12\catcode `\$12\catcode
  `\&12\catcode `\#12\catcode `\^12\catcode `\_12\catcode `\%12\relax}%
\providecommand \@@startlink[1]{}%
\providecommand \@@endlink[0]{}%
\providecommand \url  [0]{\begingroup\@sanitize@url \@url }%
\providecommand \@url [1]{\endgroup\@href {#1}{\urlprefix }}%
\providecommand \urlprefix  [0]{URL }%
\providecommand \Eprint [0]{\href }%
\providecommand \doibase [0]{http://dx.doi.org/}%
\providecommand \selectlanguage [0]{\@gobble}%
\providecommand \bibinfo  [0]{\@secondoftwo}%
\providecommand \bibfield  [0]{\@secondoftwo}%
\providecommand \translation [1]{[#1]}%
\providecommand \BibitemOpen [0]{}%
\providecommand \bibitemStop [0]{}%
\providecommand \bibitemNoStop [0]{.\EOS\space}%
\providecommand \EOS [0]{\spacefactor3000\relax}%
\providecommand \BibitemShut  [1]{\csname bibitem#1\endcsname}%
\let\auto@bib@innerbib\@empty
\bibitem [{\citenamefont {Bartholomew}(1934)}]{Bartholomew1934}%
  \BibitemOpen
  \bibfield  {author} {\bibinfo {author} {\bibfnamefont {W.~T.}\ \bibnamefont
  {Bartholomew}},\ }\href {\doibase 10.1121/1.1915685} {\bibfield  {journal}
  {\bibinfo  {journal} {The Journal of the Acoustical Society of America}\
  }\textbf {\bibinfo {volume} {6}},\ \bibinfo {pages} {25} (\bibinfo {year}
  {1934})}\BibitemShut {NoStop}%
\bibitem [{\citenamefont {Sundberg}(1990)}]{Sundberg1990}%
  \BibitemOpen
  \bibfield  {author} {\bibinfo {author} {\bibfnamefont {J.}~\bibnamefont
  {Sundberg}},\ }\href {\doibase https://doi.org/10.1016/S0892-1997(05)80135-3}
  {\bibfield  {journal} {\bibinfo  {journal} {Journal of Voice}\ }\textbf
  {\bibinfo {volume} {4}},\ \bibinfo {pages} {107 } (\bibinfo {year}
  {1990})}\BibitemShut {NoStop}%
\bibitem [{\citenamefont {Omori}\ \emph {et~al.}(1996)\citenamefont {Omori},
  \citenamefont {Kacker}, \citenamefont {Carroll}, \citenamefont {Riley},\ and\
  \citenamefont {Blaugrund}}]{Omori1996}%
  \BibitemOpen
  \bibfield  {author} {\bibinfo {author} {\bibfnamefont {K.}~\bibnamefont
  {Omori}}, \bibinfo {author} {\bibfnamefont {A.}~\bibnamefont {Kacker}},
  \bibinfo {author} {\bibfnamefont {L.~M.}\ \bibnamefont {Carroll}}, \bibinfo
  {author} {\bibfnamefont {W.~D.}\ \bibnamefont {Riley}}, \ and\ \bibinfo
  {author} {\bibfnamefont {S.~M.}\ \bibnamefont {Blaugrund}},\ }\href {\doibase
  https://doi.org/10.1016/S0892-1997(96)80003-8} {\bibfield  {journal}
  {\bibinfo  {journal} {Journal of Voice}\ }\textbf {\bibinfo {volume} {10}},\
  \bibinfo {pages} {228 } (\bibinfo {year} {1996})}\BibitemShut {NoStop}%
\bibitem [{\citenamefont {Watts}\ \emph {et~al.}(2006)\citenamefont {Watts},
  \citenamefont {Barnes-Burroughs}, \citenamefont {Estis},\ and\ \citenamefont
  {Blanton}}]{Watts2006}%
  \BibitemOpen
  \bibfield  {author} {\bibinfo {author} {\bibfnamefont {C.}~\bibnamefont
  {Watts}}, \bibinfo {author} {\bibfnamefont {K.}~\bibnamefont
  {Barnes-Burroughs}}, \bibinfo {author} {\bibfnamefont {J.}~\bibnamefont
  {Estis}}, \ and\ \bibinfo {author} {\bibfnamefont {D.}~\bibnamefont
  {Blanton}},\ }\href@noop {} {\bibfield  {journal} {\bibinfo  {journal}
  {Journal of voice}\ }\textbf {\bibinfo {volume} {20}},\ \bibinfo {pages} {82}
  (\bibinfo {year} {2006})}\BibitemShut {NoStop}%
\bibitem [{\citenamefont {Aithal}\ \emph {et~al.}(2011)\citenamefont {Aithal},
  \citenamefont {Swathi},\ and\ \citenamefont {Rajasudhakar}}]{Aithal2011}%
  \BibitemOpen
  \bibfield  {author} {\bibinfo {author} {\bibfnamefont {S.}~\bibnamefont
  {Aithal}}, \bibinfo {author} {\bibfnamefont {S.}~\bibnamefont {Swathi}}, \
  and\ \bibinfo {author} {\bibfnamefont {R.}~\bibnamefont {Rajasudhakar}},\
  }\href@noop {} {\bibfield  {journal} {\bibinfo  {journal} {ninad}\ }\textbf
  {\bibinfo {volume} {25}},\ \bibinfo {pages} {46} (\bibinfo {year}
  {2011})}\BibitemShut {NoStop}%
\bibitem [{\citenamefont {Usha}\ \emph {et~al.}(2017)\citenamefont {Usha},
  \citenamefont {Geetha},\ and\ \citenamefont {Darshan}}]{Usha2017}%
  \BibitemOpen
  \bibfield  {author} {\bibinfo {author} {\bibfnamefont {M.}~\bibnamefont
  {Usha}}, \bibinfo {author} {\bibfnamefont {Y.}~\bibnamefont {Geetha}}, \ and\
  \bibinfo {author} {\bibfnamefont {Y.}~\bibnamefont {Darshan}},\ }\href@noop
  {} {\bibfield  {journal} {\bibinfo  {journal} {Journal of Voice}\ }\textbf
  {\bibinfo {volume} {31}},\ \bibinfo {pages} {157} (\bibinfo {year}
  {2017})}\BibitemShut {NoStop}%
\bibitem [{\citenamefont {Manfredi}\ \emph {et~al.}(2015)\citenamefont
  {Manfredi}, \citenamefont {Barbagallo}, \citenamefont {Baracca},
  \citenamefont {Orlandi}, \citenamefont {Bandini},\ and\ \citenamefont
  {Dejonckere}}]{Manfredi2015}%
  \BibitemOpen
  \bibfield  {author} {\bibinfo {author} {\bibfnamefont {C.}~\bibnamefont
  {Manfredi}}, \bibinfo {author} {\bibfnamefont {D.}~\bibnamefont
  {Barbagallo}}, \bibinfo {author} {\bibfnamefont {G.}~\bibnamefont {Baracca}},
  \bibinfo {author} {\bibfnamefont {S.}~\bibnamefont {Orlandi}}, \bibinfo
  {author} {\bibfnamefont {A.}~\bibnamefont {Bandini}}, \ and\ \bibinfo
  {author} {\bibfnamefont {P.~H.}\ \bibnamefont {Dejonckere}},\ }\href
  {\doibase https://doi.org/10.1016/j.jvoice.2014.09.014} {\bibfield  {journal}
  {\bibinfo  {journal} {Journal of Voice}\ }\textbf {\bibinfo {volume} {29}},\
  \bibinfo {pages} {517.e1 } (\bibinfo {year} {2015})}\BibitemShut {NoStop}%
\bibitem [{\citenamefont {Welch}\ \emph {et~al.}(1989)\citenamefont {Welch},
  \citenamefont {Howard},\ and\ \citenamefont {Rush}}]{Welch1989}%
  \BibitemOpen
  \bibfield  {author} {\bibinfo {author} {\bibfnamefont {G.~F.}\ \bibnamefont
  {Welch}}, \bibinfo {author} {\bibfnamefont {D.~M.}\ \bibnamefont {Howard}}, \
  and\ \bibinfo {author} {\bibfnamefont {C.}~\bibnamefont {Rush}},\ }\href@noop
  {} {\bibfield  {journal} {\bibinfo  {journal} {Psychology of Music}\ }\textbf
  {\bibinfo {volume} {17}},\ \bibinfo {pages} {146} (\bibinfo {year}
  {1989})}\BibitemShut {NoStop}%
\bibitem [{\citenamefont {Hoppe}\ \emph {et~al.}(2006)\citenamefont {Hoppe},
  \citenamefont {Sadakata},\ and\ \citenamefont {Desain}}]{Hoppe2006}%
  \BibitemOpen
  \bibfield  {author} {\bibinfo {author} {\bibfnamefont {D.}~\bibnamefont
  {Hoppe}}, \bibinfo {author} {\bibfnamefont {M.}~\bibnamefont {Sadakata}}, \
  and\ \bibinfo {author} {\bibfnamefont {P.}~\bibnamefont {Desain}},\
  }\href@noop {} {\bibfield  {journal} {\bibinfo  {journal} {Journal of
  computer assisted learning}\ }\textbf {\bibinfo {volume} {22}},\ \bibinfo
  {pages} {308} (\bibinfo {year} {2006})}\BibitemShut {NoStop}%
\bibitem [{\citenamefont {{Tsai}}\ and\ \citenamefont
  {{Lee}}(2012)}]{Tsai2012}%
  \BibitemOpen
  \bibfield  {author} {\bibinfo {author} {\bibfnamefont {W.}~\bibnamefont
  {{Tsai}}}\ and\ \bibinfo {author} {\bibfnamefont {H.}~\bibnamefont {{Lee}}},\
  }\href {\doibase 10.1109/TASL.2011.2174224} {\bibfield  {journal} {\bibinfo
  {journal} {IEEE Transactions on Audio, Speech, and Language Processing}\
  }\textbf {\bibinfo {volume} {20}},\ \bibinfo {pages} {1233} (\bibinfo {year}
  {2012})}\BibitemShut {NoStop}%
\bibitem [{\citenamefont {Goto}(2014)}]{Goto2014}%
  \BibitemOpen
  \bibfield  {author} {\bibinfo {author} {\bibfnamefont {M.}~\bibnamefont
  {Goto}},\ }in\ \href@noop {} {\emph {\bibinfo {booktitle} {2014 12th
  International Conference on Signal Processing (ICSP)}}}\ (\bibinfo
  {organization} {IEEE},\ \bibinfo {year} {2014})\ pp.\ \bibinfo {pages}
  {2431--2438}\BibitemShut {NoStop}%
\bibitem [{\citenamefont {Nakano}\ \emph {et~al.}(2006)\citenamefont {Nakano},
  \citenamefont {Goto},\ and\ \citenamefont {Hiraga}}]{Nakano2006}%
  \BibitemOpen
  \bibfield  {author} {\bibinfo {author} {\bibfnamefont {T.}~\bibnamefont
  {Nakano}}, \bibinfo {author} {\bibfnamefont {M.}~\bibnamefont {Goto}}, \ and\
  \bibinfo {author} {\bibfnamefont {Y.}~\bibnamefont {Hiraga}},\ }in\
  \href@noop {} {\emph {\bibinfo {booktitle} {Ninth International Conference on
  Spoken Language Processing}}}\ (\bibinfo {year} {2006})\BibitemShut {NoStop}%
\bibitem [{\citenamefont {Furukawa}\ and\ \citenamefont
  {Shinomoto}(2018)}]{Furukawa2018}%
  \BibitemOpen
  \bibfield  {author} {\bibinfo {author} {\bibfnamefont {M.}~\bibnamefont
  {Furukawa}}\ and\ \bibinfo {author} {\bibfnamefont {S.}~\bibnamefont
  {Shinomoto}},\ }\href@noop {} {\bibfield  {journal} {\bibinfo  {journal}
  {Neural Computing and Applications}\ }\textbf {\bibinfo {volume} {30}},\
  \bibinfo {pages} {2471} (\bibinfo {year} {2018})}\BibitemShut {NoStop}%
\bibitem [{\citenamefont {Bishop}(2006)}]{Bishop2006}%
  \BibitemOpen
  \bibfield  {author} {\bibinfo {author} {\bibfnamefont {C.~M.}\ \bibnamefont
  {Bishop}},\ }\href@noop {} {\emph {\bibinfo {title} {Pattern recognition and
  machine learning}}}\ (\bibinfo  {publisher} {springer},\ \bibinfo {year}
  {2006})\BibitemShut {NoStop}%
\bibitem [{\citenamefont {Murphy}(2012)}]{Murphy2012}%
  \BibitemOpen
  \bibfield  {author} {\bibinfo {author} {\bibfnamefont {K.~P.}\ \bibnamefont
  {Murphy}},\ }\href@noop {} {\emph {\bibinfo {title} {Machine learning: a
  probabilistic perspective}}}\ (\bibinfo  {publisher} {MIT press},\ \bibinfo
  {year} {2012})\BibitemShut {NoStop}%
\bibitem [{\citenamefont {B{\"o}hm}\ \emph {et~al.}(2017)\citenamefont
  {B{\"o}hm}, \citenamefont {Eyben}, \citenamefont {Schmitt}, \citenamefont
  {Kosch},\ and\ \citenamefont {Schuller}}]{Bohm2017}%
  \BibitemOpen
  \bibfield  {author} {\bibinfo {author} {\bibfnamefont {J.}~\bibnamefont
  {B{\"o}hm}}, \bibinfo {author} {\bibfnamefont {F.}~\bibnamefont {Eyben}},
  \bibinfo {author} {\bibfnamefont {M.}~\bibnamefont {Schmitt}}, \bibinfo
  {author} {\bibfnamefont {H.}~\bibnamefont {Kosch}}, \ and\ \bibinfo {author}
  {\bibfnamefont {B.}~\bibnamefont {Schuller}},\ }in\ \href@noop {} {\emph
  {\bibinfo {booktitle} {2017 International Joint Conference on Neural Networks
  (IJCNN)}}}\ (\bibinfo {organization} {IEEE},\ \bibinfo {year} {2017})\ pp.\
  \bibinfo {pages} {1560--1569}\BibitemShut {NoStop}%
\bibitem [{\citenamefont {Vidwans}\ \emph {et~al.}(2017)\citenamefont
  {Vidwans}, \citenamefont {Gururani}, \citenamefont {Wu}, \citenamefont
  {Subramanian}, \citenamefont {Swaminathan},\ and\ \citenamefont
  {Lerch}}]{Vidwans2017}%
  \BibitemOpen
  \bibfield  {author} {\bibinfo {author} {\bibfnamefont {A.}~\bibnamefont
  {Vidwans}}, \bibinfo {author} {\bibfnamefont {S.}~\bibnamefont {Gururani}},
  \bibinfo {author} {\bibfnamefont {C.-W.}\ \bibnamefont {Wu}}, \bibinfo
  {author} {\bibfnamefont {V.}~\bibnamefont {Subramanian}}, \bibinfo {author}
  {\bibfnamefont {R.~V.}\ \bibnamefont {Swaminathan}}, \ and\ \bibinfo {author}
  {\bibfnamefont {A.}~\bibnamefont {Lerch}},\ }in\ \href@noop {} {\emph
  {\bibinfo {booktitle} {Audio Engineering Society Conference: 2017 AES
  International Conference on Semantic Audio}}}\ (\bibinfo {organization}
  {Audio Engineering Society},\ \bibinfo {year} {2017})\BibitemShut {NoStop}%
\bibitem [{\citenamefont {Pati}\ \emph {et~al.}(2018)\citenamefont {Pati},
  \citenamefont {Gururani},\ and\ \citenamefont {Lerch}}]{Pati2018}%
  \BibitemOpen
  \bibfield  {author} {\bibinfo {author} {\bibfnamefont {K.}~\bibnamefont
  {Pati}}, \bibinfo {author} {\bibfnamefont {S.}~\bibnamefont {Gururani}}, \
  and\ \bibinfo {author} {\bibfnamefont {A.}~\bibnamefont {Lerch}},\
  }\href@noop {} {\bibfield  {journal} {\bibinfo  {journal} {Applied Sciences}\
  }\textbf {\bibinfo {volume} {8}},\ \bibinfo {pages} {507} (\bibinfo {year}
  {2018})}\BibitemShut {NoStop}%
\bibitem [{\citenamefont {Wu}\ and\ \citenamefont {Lerch}(2018)}]{Wu2018}%
  \BibitemOpen
  \bibfield  {author} {\bibinfo {author} {\bibfnamefont {C.-W.}\ \bibnamefont
  {Wu}}\ and\ \bibinfo {author} {\bibfnamefont {A.}~\bibnamefont {Lerch}},\
  }in\ \href@noop {} {\emph {\bibinfo {booktitle} {2018 IEEE 12th International
  Conference on Semantic Computing (ICSC)}}}\ (\bibinfo {organization} {IEEE},\
  \bibinfo {year} {2018})\ pp.\ \bibinfo {pages} {93--99}\BibitemShut {NoStop}%
\bibitem [{\citenamefont {Zhang}\ \emph {et~al.}(2019)\citenamefont {Zhang},
  \citenamefont {Jiang}, \citenamefont {Deng},\ and\ \citenamefont
  {Li}}]{Zhang2019}%
  \BibitemOpen
  \bibfield  {author} {\bibinfo {author} {\bibfnamefont {N.}~\bibnamefont
  {Zhang}}, \bibinfo {author} {\bibfnamefont {T.}~\bibnamefont {Jiang}},
  \bibinfo {author} {\bibfnamefont {F.}~\bibnamefont {Deng}}, \ and\ \bibinfo
  {author} {\bibfnamefont {Y.}~\bibnamefont {Li}},\ }in\ \href@noop {} {\emph
  {\bibinfo {booktitle} {ICASSP 2019-2019 IEEE International Conference on
  Acoustics, Speech and Signal Processing (ICASSP)}}}\ (\bibinfo {organization}
  {IEEE},\ \bibinfo {year} {2019})\ pp.\ \bibinfo {pages}
  {466--470}\BibitemShut {NoStop}%
\end{thebibliography}

%

\end{document}